\documentstyle[twoside,fleqn,espcrc2]{article}

\def\gsim{\stackrel{>}{\sim}}
\def\lsim{\stackrel{<}{\sim}}
\def\beq{\begin{equation}}
\def\eeq{\end{equation}}

\def\pbar{\bar{p}}

\newcommand{\AmS}{{\protect\the\textfont2
  A\kern-.1667em\lower.5ex\hbox{M}\kern-.125emS}}

\title{c\thanks
      {This work was supported in part by the U.S. Department of Energy 
                grant no. DE-FG05-85ER40226.}}

\author{Thomas J. Weiler%
        \thanks{Speaker at DM96, UCLA, Febr. 14--16, 1996.}
        and Thomas W. Kephart\\ \medskip
                  Department of Physics \& Astronomy, Vanderbilt University,
	        Nashville, TN 37235}

\begin{document}

\begin{abstract}
We suggest that the highest energy $ \gsim 10^{20}$ eV
cosmic ray primaries may be relativistic magnetic monopoles.
Motivations for this hypothesis are twofold: 
(i) conventional primaries are problematic, while 
monopoles are naturally accelerated to $E \sim 10^{20}$ eV
by galactic magnetic fields;
(ii) the observed highest energy cosmic ray flux is just below the 
Parker limit for monopoles.
By matching the cosmic monopole production mechanism to the observed 
highest energy cosmic ray flux
we estimate the monopole mass to be $\lsim 10^{10}$ GeV.
\end{abstract}

\maketitle

The recent discoveries by the AGASA \cite{akeno}, Fly's Eye \cite{eye}, 
Haverah Park \cite{hp}, and Yakutsk \cite{yak} collaborations of cosmic rays with 
energies above the GZK \cite{gzk} cut--off at $E_c\sim 5\times10^{19}$ eV present
an intriguing challenge to particle astrophysics.  
The origin of the cut--off is degradation of the proton energy by 
resonant scattering 
on the $3K$ cosmic background radiation; above threshold, a $\Delta^*$ 
is produced which then decays to nucleon plus pion.
For every mean free path $\sim 6$ Mpc of travel,
the proton loses 20\% of its energy on average.
So if protons are the primaries for the 
highest energy cosmic rays they must either come from a rather nearby source
($\lsim 50$ to 100 Mpc \cite{sommers})
or have an initial energy far above $10^{20}$ eV. 
Neither possibility seems likely, although the suggestion has been made
that radio galaxies at distances 10 to $200\,h_{100}^{-1}$ Mpc 
in the supergalactic plane 
may be origins \cite{biermann}.
A primary nucleus mitigates the cut--off problem (energy per nucleon
is reduced by 1/A), but has additional problems:
above $\sim 10^{19}$ eV nuclei
should be photo--dissociated by the 3K background \cite{stecker}, 
and possibly disintegrated by the 
particle density ambient at the astrophysical source.  

Gamma--rays and neutrinos are other possible primary candidates 
for these highest energy events.
However, the gamma--ray hypothesis appears inconsistent \cite{hs}
with the time--development of the Fly's Eye event.
In addition, the mean free path for a $\sim 10^{20}$ eV photon to annihilate
on the radio background to $e^+ e^-$ is believed to be only
$10$ to $40$ Mpc \cite{hs}, and the density profile of the Yakutsk event \cite{yak}
showed a large number of muons which argues against gamma--ray initiation.
Concerning the neutrino hypothesis, 
the Fly's Eye event occured high in the atmosphere,
whereas the expected event rate for early development of a neutrino--induced 
air shower is down from that of an 
electromagnetic or hadronic interaction by six orders of magnitude \cite{hs}.
Moreover, the acceleration problem for $\gamma$ and $\nu$ primaries is as 
daunting as for hadrons, 
since $\gamma$'s and $\nu$'s at these energies 
are believed to originate in decay of $\gsim10^{20}$ eV pions.

Given the problems with interpreting the highest energy cosmic ray 
primaries as protons, nuclei, photons, or neutrinos, 
we rekindle the idea \cite{porter} that the primary particles of the 
highest energy cosmic rays may be magnetic monopoles \cite{kw96}.
Two ``coincidences'' in the data support this hypothesis.
The first is that the energies above the cut-off are naturally
attained by monopoles when accelerated by known cosmic magnetic fields.  
The second is that the observed cosmic ray flux above the cut--off is of the
same order of magnitude as the theoretically allowed ``Parker limit'' 
monopole flux.

To impart its kinetic energy to the induced air--shower,
the monopole must be relativistic.  This bounds the monopole mass 
to be $\lsim 10^{10}$ GeV.  
The Kibble mechanism \cite{kibble} for monopole generation in 
an early--universe phase transition establishes a monotonic relationship
between the monopole's flux and mass.  There results, then,
a second upper bound on the monopole mass,
which turns out to be similar.
The consistency of these two bounds is a third ``coincidence.''
Thus, we arrive at a flux of monopoles of mass $M \lsim 10^{10}$ GeV as
a viable explanation the highest energy cosmic ray data.
This hypothesis has testable signatures, as we shall see.

The kinetic energy of cosmic monopoles is easily obtained.
As pointed out by Dirac, the minimum charge for a monopole is 
$q_M = e/{2\alpha}$ (which implies $\alpha_M =1/4\alpha$).
In the local interstellar medium, the magnetic field $B$ is 
approximately $3 \times10^{-6}$ gauss ($\equiv B_{-6}$) with a coherence 
length $L\sim 300$ pc ($\equiv L_{300}$) \cite{kt}. 
Thus, a galactic monopole will typically have kinetic energy:
\begin{eqnarray*}
E_K & \sim q_M B L \sqrt{N} & \\ 
    & \simeq 6\times 10^{20} & \!\!\!\!(\frac{B}{B_{-6}}) \;
										(\frac{L}{L_{300}})^{1/2}\;(\frac{R_M}{R_{30}})^{1/2}\;{\rm eV},
\end{eqnarray*}
where $N\sim R_M/L \sim 100\,(R_M/R_{30})/(L/L_{300})$ 
is the number of magnetic domains encountered by a typical monopole as it 
traverses the galactic magnetic field region of size 
$R_M\equiv R_{30} \times 30$ kpc.
Note that this energy is above the GZK cut--off.
Thus, the ``acceleration problem'' for $E\gsim 10^{20}$ eV primaries 
is naturally solved in the monopole hypothesis.

Another monopole acceleration mechanism of the right order of magnitude 
is provided by the surface magnetic field of a neutron star.
At the neutron star's surface, a monopole acquires a kinetic energy
$E_K\equiv q_M B L \simeq 
2\times10^{21} {\rm eV} (B/10^{12} {\rm gauss}) (L/{\rm km}).$
However, it is thought to be unlikely that objects as small as stars would
contain a population of bound monopoles large enough to generate a 
measureable flux.

To obtain the theoretically predicted monopole flux,
it is worthwhile to review
how and when a monopole is generated in a phase transition \cite{kibble,kt}.
The topological requirement for monopole production 
is that a semisimple gauge group changes so that
a $U(1)$ factor becomes unbroken.  
If the mass or temperature scale at which the symmetry changes is $\Lambda$, 
then the monopoles appear as topological defects, with 
mass $M\sim \alpha^{-1} \Lambda$.
We use $M\sim 100\,\Lambda$ in the estimates to follow.
All that is necessary to ensure that the monopoles are relativistic today, 
i.e. $M \lsim 10^{10}$ GeV, and so produce relativistic air showers,
is to require this symmetry breaking scale associated 
with the production of monopoles to be at or below  $\sim 10^8$ GeV.

This $M\lsim 10^{10}$ GeV restriction also serves to ameliorate possible 
overclosure of the universe by an excessive monopole mass density.
At the time of the phase transition, roughly one monopole 
or antimonopole is produced per correlated volume \cite{kibble}. 
The resulting monopole number density today is 
\beq
n_M \sim 0.1\, (\Lambda/10^{17} {\rm GeV})^3 (l_H/\xi_c)^3 {\rm cm}^{-3},
\label{density}
\eeq
where $\xi_c$ is the phase transition correlation length, 
bounded from above
by the horizon size $l_H$ at the time of the phase transition, or
equivalently, at the Ginsburg temperature $T_G$ of the phase transition.
The correlation length may be comparable to the horizon size
(second order or weakly first order phase transition)
or considerably smaller than the horizon size
(strongly first order transition).
The resulting monopole mass density today relative to the closure value is
\beq
\Omega_M \sim 0.1\, (M/10^{13} {\rm GeV})^4 (l_H/\xi_c)^3.
\label{omega}
\eeq
Monopoles less massive than 
$\sim 10^{13} (\xi_c/l_H)^{3/4}$ GeV do not overclose the universe.

From Eq.(\ref{density}),
the general expression for the relativistic monopole 
flux may be written
\beq
F_M = c\: n_M/4\pi
 \sim 0.2\, (M /{10^{16} {\rm GeV}})^3(l_H/\xi_c)^3
\label{flux}
\eeq
per ${\rm cm}^2 \cdot$sec$\cdot$sr.
The ``Parker limit'' on the galactic monopole flux \cite{parker} is
$F_M^{PL}\leq 10^{-15}/{\rm cm}^2/{\rm sec}/{\rm sr}$. 
It is derived by requiring that the measured galactic magnetic fields 
not be depleted (by accelerating monopoles) faster than the fields can be regenerated 
by galactic magnetohydrodynamics.
Comparing this Parker limit with the general monopole flux in Eq. (\ref{flux}),
we see that the Parker bound is satisfied if 
$M\lsim 10^{11} (\xi_c/l_H)$ GeV.  
From Eqs. (\ref{omega}) and (\ref{flux}) we may also write for the {\it relativistic}
monopole closure density
$\Omega_{RM} \sim 10^{-8} (\langle E_M \rangle /10^{20} {\rm eV}) (F_M/F_M^{PL})$,
which shows that the hypothesized monopole flux does not close the universe
regardless of the nature of the monopole--creating 
phase transition (parameterized by $\xi_c/l_H$).

There is no obvious reason why monopoles accelerated by cosmic magnetic fields 
should have a falling spectrum, or even a broad spectrum.
So we assume that the monopole spectrum is peaked
in the energy half--decade 1 to $5\times 10^{20}$ eV.
With this assumption, the monopole differential flux is
\[
\frac{dF_M}{dE}\sim 4\times 10^{-40} (\frac{M}{10^{10} {\rm GeV}})^3
    (\frac{l_H}{\xi_c})^3
\]
per ${\rm cm}^2 \cdot$sec$\cdot$sr$\cdot$eV.
Comparing this monopole flux to the measured differential flux
$(dF/dE)_{Exp} \sim 10^{-38 \pm 2}$ 
per ${\rm cm}^2 \cdot$sec$\cdot$sr$\cdot$eV
above $10^{20}$ eV (summarized in \cite{hs}),
we infer $M \sim (\xi_c/l_H)\times 10^{10 \pm 1}$ GeV.
We note that the monopole mass derived here from the flux requirement is 
remarkably consistent with the three prior mass requirements,
namely that the $E\sim 10^{20}$ eV monopoles be relativistic, 
that they not overclose the universe, and that they obey the Parker limit. 
It is very interesting that the observed highest energy cosmic ray flux
lies just below the Parker limit for monopole flux.  
A slightly larger observed flux would exceed this limit, 
while a slightly lower flux would not have been observed.
If the monopole hypothesis is correct,
it is possible that we are seeing evidence for some 
dynamical reason forcing the monopole flux 
to saturate the Parker bound.

Let us analyze the monopole hypothesis in detail by focussing on 
some more salient features of the data.
There appears to be an event pile--up just below $\sim 6\times 10^{19}$ eV
(the GZK cut--off), and a gap just above.
There are events above the gap which we propose to explain.
So far, no events are seen above the Fly's Eye event energy at 
$3\times 10^{20}$ eV.
The event rate at highest energies exceeds
a power law extrapolation from the spectrum below the gap 
(with low statistical significance).
Except for the highest energy cosmic ray events, 
the spectrum is well fit \cite{akeno} by a diffuse population of
protons distributed isotropically in the universe.
The apparent pile--up of events between 
$\sim 10^{19}$ eV and $6 \times10^{19}$ eV is explained by the
pion photo--production mechanism of GZK \cite{pileup}. 
For the events above $10^{20}$ eV, a different origin seems to be required.
That the galactic magnetic fields naturally impart 
$10^{20}$ to $10^{21}$ eV of kinetic energy to the monopole, and that 
there appears to be an absence of events above and just below this energy,
we find very suggestive.
A monopole with $\gamma_M \equiv E_M/M$ will 
forward--scatter atmospheric particles to $\gamma=2 \gamma_M^2$.
Consequently, there is an effective energy threshold of $E_M\sim 10\,M$ for 
relativistic air showers induced by monopoles.
Thus, an apparent threshold in the data at $E\sim 10^{20}$ eV
may also be explained if the monopole mass is $\sim 10^{10}$ GeV.

Any proposed primary candidate must be able to reproduce 
the observed shower evolution of the $3\times 10^{20}$ eV Fly's Eye event. 
The shower peaks at $815\pm 55 \; {\rm g/cm}^2$, which is marginally consistent with
that expected in a proton--initiated shower.
Does a monopole--induced air shower fit the Fly's Eye event profile?
We do not know.  
The hadronic component of the monopole shower is 
likely to be complicated.   
The interior of the monopole is symmetric vacuum, in which
all the fermion, Yang--Mills, and Higgs fields of the grand unified theory
coexist.  Thus, even though the Compton size of the monopole is
incredibly tiny, its strong interaction size is the usual confinement
radius of $\sim$ 1 fm, and its strong interaction cross--section is
indeed strong, $\sim 10^{-26} {\rm cm}^2$, and possibly growing
with energy like other hadronic cross--sections. 
Furthermore, a number of unusual monopole--nucleus interactions may take place,
including enhanced monopole--catalyzed baryon--violating processes
with a strong cross--section $\sim 10^{-27} {\rm cm}^2$ \cite{CallanRub};
catalyzation of the inverse 
process $e^- + M  \rightarrow M + \pi + (\pbar$ or ${\bar n})$, followed by
pion/antibaryon initiation of a hadronic shower;
binding of one or more nucleons by the monopole \cite{craigie}, 
in which case the monopole--air interaction may resemble a
a relativistic nucleus--nucleus collision;
strong polarization of the air nuclei due to magnetic
interaction with the individual nucleon magnetic moments 
and electric ($\vec{E} =\gamma_M e/2\alpha r^2 \hat{\phi}$) 
interaction with the proton constituents,
possibly causing fragmentation \cite{craigie};
hard elastic magnetic scattering of ionized nuclei 
(in the rest frame of the monopole the charged nucleus 
will see the monopole as a reflecting magnetic mirror);
and possible electroweak--scale sphaleron processes \cite{sphaleron}
at the large Q--value of the monopole--air nucleus interaction
($\sim \gamma_M A m_N \sim$ TeV).
Clearly, more theoretical work
is required to understand a monopole's air shower development.

On the other hand, 
the monopole's electromagnetic showering properties are straightforward.
A magnetic monopole has a rest--frame magnetic field $B_{RF}=q_M\hat{r}/r^2$.
When boosted to a velocity $\vec{\beta}_M$, an electric field 
$\vec{E}_M=\gamma_M\, \vec{\beta}_M \times\vec{B}_{RF}$ is generated, 
leading to a ``dual Lorentz'' force  acting on the charged 
constituents of air atoms.  The
electromagnetic energy loss of a relativistic monopole traveling through 
matter is very similar to that of a heavy 
nucleus with similar $\gamma$--factor and 
charge $Z= q_M/e =1/2\alpha =137/2$. 
One result is a 
$\sim 6 \, {\rm GeV}/({\rm g\,cm}^{-2})$ ``minimum--ionizing monopole" electromagnetic 
energy loss.
Integrated through the atmosphere, 
the total electromagnetic energy loss is therefore 
$\sim (6.2/\cos\theta_z)$ TeV, for zenith angle $\theta_z\lsim 60^{\circ}$.
For a horizontal shower the
integrated energy loss is $\sim 240$ TeV. 
A second electromagnetic prediction 
is Cerenkov radiation at the usual angle but
enhanced by $(137/2)^2 \sim 4700$ compared to a proton primary.
This enhanced Cerenkov radiation may help in the identification of the
monopole primary.

We can derive useful information on some of the characteristics of
the monopole shower simply from kinematics.
For relativistic monopoles with mass $M$ greatly exceeding 
the masses of the target air atoms
and their constituent nucleon masses $m$, 
the maximum energy transfer occurs via forward (in the lab frame) 
elastic scattering.  This maximum is 
\[
E^{'}_m/E_M = (1+M^2/2mE_M)^{-1}.
\]
In contrast, the maximum energy transfer for a relativistic
particle of energy $E$ and mass $m$ scattering on a stationary target particle 
of the same mass is
\[
E^{'}_m/E_m = 1-m/2E \sim 1.
\]
We see that a relativistic nucleon or 
light nucleus primary will transfer essentially all of its energy in a single
forward scattering event.  
If the monopole has $M\lsim \sqrt{2m E_M}$, i.e. $\lsim 10^6$ GeV 
for $E_M\sim {\rm few}\times10^{20}$ eV,
it too will transfer most of its energy in the first forward--scattering event, 
possibly mimicking a standard air shower.
On the other hand,
a relativistic monopole primary with $M > 10^6$ GeV will retain most of its energy 
per each scattering, and so will continuously ``initiate'' the shower as it 
propagates through the atmosphere.
For this reason, we refer to the monopole shower as ``monopole--induced'' rather 
than ``monopole--initiated.'' 
The smaller energy transfer per collision for a $M >10^6$ GeV monopole as compared
to that of the usual primary candidates may constitute a signature for heavy 
monopole primaries.  Moreover, the back--scattered atmospheric particles in the 
center--of--mass system (which is roughly half of the scattered particles) are 
forward--scattered in the lab frame into a cone of half--angle $1/\gamma_M$;
at the given energy of $E\sim 10^{20}$ eV, 
this angle will be large for a heavy monopole primary compared to the angle 
for a usual primary particle, possibly offering another monopole signature.  

Simple GUT models may be constructed in which a $U(1)$ symmetry first appears
at a cosmic temperature 
far below the initial GUT--breaking scale, signaling the appearance of monopoles 
with mass $M$ far below the initial GUT scale. 
Indeed, there are several published models in which exactly this happens,
the most recent being \cite{desh96}.
The utility of an intermediate breaking scale has been 
invoked before in many contexts, including
the Peccei--Quinn solution to the strong CP problem,
the right--handed neutrino scale in ``see--saw'' models
of neutrino mass generation, and supersymmetry breaking in a hidden sector. 

To conclude, we suggest that the primary particles of 
the highest energy cosmic rays discovered in the past several 
years are relativistic magnetic monopoles of mass $M\lsim 10^{10}$ GeV.
Energies of $\sim 10^{20}$ eV can easily be attained via
acceleration in a typical galactic magnetic field, and the observed
highest energy cosmic ray flux (just below the Parker limit) can be explained within 
the monopole hypothesis by the Kibble mechanism.
Fortunately, there are some possible tests of this monopole hypothesis.
First of all, the monopole primaries should be asymmetrically 
distributed on the sky, showing a preference for the direction 
of the local galactic magnetic field.
Secondly, the characteristics of air showers induced by 
monopoles may carry distinctive signatures:  
The electromagnetic shower and Cerenkov cone should develop as if 
the relativistic monopole carried 
$\sim 137/2$ units of electric charge. 
In addition, 
there may be several strong interaction aspects of the monopole, each 
contributing to monopole--induced air shower development.
Finally, the energy transfer per scatterer will be smaller for a 
$M\gsim 10^6$ GeV monopole compared to that of a standard primary,
and the scattering angle will be larger.

There are good prospects for more cosmic ray data at these 
highest energies.
The present cosmic ray 
detection efforts are ongoing, and the ``Auger Project''
has been formed to coordinate an international 
effort to instrument a 5,000 ${\rm km}^2$ detector and
collect five thousand events per year above $10^{19}$ \cite{cronin}.


\begin{thebibliography}{99}

\bibitem{akeno} S.Yoshida, et al., (AGASA Collab.) 
   {\sl Astropart., Phys.} {\bf 3}, 105 (1995);
    N. Hayashida et al., {\sl Phys. Rev. Lett.} {\bf 73}, 3491 (1994).

\bibitem{eye} D. J. Bird et al., (Fly's Eye Collab.) 
   {\sl  Phys. Rev. Lett.} {\bf 71}, 3401 (1993);
   {\sl Astrophys. J.} {\bf 424}, 491 (1994); 
   {\sl ibid.} {\bf 441}, 144 (1995).
  
\bibitem{hp} G. Brooke et al. (Haverah Park Collab.), 
   Proc. 19th Intl. Cosmic Ray Conf. (La Jolla) {\bf 2}, 150 (1985);
   reported in M. A. Lawrence, R. J. O. Reid, and A. A. Watson
    (Haverah Park Collab.), 
   {\sl J. Phys. G} {\bf 17}, 733 (1991).

\bibitem{yak} N. N. Efimov et al., (Yakutsk Collab.) 
   ICRR Symposium on Astrophysical Aspects  
   of the Most Energetic Cosmic  Rays, ed. N. Nagano and F. Takahara,
   World Scientific pub. (1991);
   and Proc. 22nd ICRC, Dublin (1991).

\bibitem{gzk} K. Greisen,{\sl Phys. Rev. Lett.} {\bf 16}, 748 (1966);
   G. T. Zatsepin and V. A. Kuzmin,
   {\sl Pisma Zh. Eksp. Teor. Fiz.} {\bf 4}, 114 (1966);
    J. L. Puget, F. W. Stecker and J. H. Bredekamp, {\sl Ap. J.}, {\bf 205}, 
    638 (1976);
    V. S. Berezinsky and S. I. Grigoreva, {\sl Astron. Astrophys.}, 
    {\bf199}, 1 (1988);
    S. Yoshida and M. Teshima, {\sl Prog. Theor. Phys.}, {\bf 89}, 833 (1993).

\bibitem{sommers} J. W. Elbert and P. Sommers, 
   {\sl Astrophys. J.} {\bf 441}, 151 (1995);
   G. Sigl, D. N. Schramm, 
   and P. Bhattacharjee,
   {\sl Astropart. Phys.} {\bf 2}, 401 (1994).

\bibitem{biermann}  T.Stanev, P. Biermann, J. Lloyd-Evans, J. Rachen and
  A. Watson, {\sl Phys. Rev. Lett.} {\bf 75}, 3056 (1995).

\bibitem{stecker} F. W. Stecker, {\sl Phys. Rev.} {\bf 180}, 1264 (1969).

\bibitem{hs} F. Halzen, R. A. Vazquez, T. Stanev, and V. P Vankov, 
  {\sl Astropart., Phys.}, {\bf 3}, 151 (1995).

\bibitem{porter} N. A. Porter, {\sl Nuovo Cim.} {\bf 16}, 958 (1960).

\bibitem{kw96} T. W. Kephart and T. J. Weiler, 
					{\sl Astropart. Phys.} {\bf 4}, 271 (1996).

\bibitem{kibble} T. W. B. Kibble, {\sl J. Phys.} {\bf A9}, 1387 (1976),
   and {\sl Phys. Rept.} {\bf 67}, 183 (1980); 
   M. B. Einhorn, D. L. Stein, and D. Toussaint, 
   {\sl Phys. Rev.} {\bf D21}, 3295 (1980); 
   A. H. Guth and E. J. Weinberg, {\sl Phys. Rev.} {\bf D23}, 876 (1981).

\bibitem{kt} E. W. Kolb and M. S. Turner, ``The Early Universe," 
				Addison-Wesley pub., NY (1991).

\bibitem{parker} E. N. Parker, {\sl Astrophys. J.} {\bf 160}, 383 (1970);
	  {\it ibid.\ }, {\bf 163}, 225 (1971); {\it ibid.\ },{\bf 166}, 295 (1971);
   M. S. Turner, E. N. Parker, and T. Bogdan,
   {\sl Phys. Rev.} {\bf D26}, 1296 (1982).

\bibitem{pileup} C. T. Hill and D. N. Schramm, 
   {\sl  Phys. Rev.} {\bf D31 }, 564 (1985);
   F. A. Aharonian and J. W. Cronin,
   {\sl Phys. Rev.} {\bf D50}, 1892 (1994).

\bibitem{CallanRub} V. Rubakov, {\sl JETP Lett.} {\bf 33}, 644 (1981);
   {\it Nucl. Phys.} {\bf B203}, 311 (1982);
   C. G. Callan, Jr., {\sl Phys. Rev.} {\bf 25}, 2141 (1982).

\bibitem{craigie} G. Giacomelli, in ``Theory and Detection of Magnetic Monopoles
  in Gauge Theories'', ed. N. Craigie, World Scientific pub., 1986;
   K. Olaussen, H. A. Olsen, P. Osland, and I. Overbo,
   {\sl Phys. Rev. Lett.} {\bf 52}, 325 (1984);

\bibitem{sphaleron} A sphaleron is the minimum--energy baryon-- and 
   lepton--number violating classical field configuration of the standard model.
   An overview of sphaleron physics can be found 
   in ref. \cite{kt}.

\bibitem{desh96} N. G. Deshpande, B. Dutta, and E. Keith, 
   net preprint hep-ph/9604236.

\bibitem{cronin} J. W. Cronin and A. A. Watson, announcement from the 
   Giant Air Shower Design Group (recently renamed the ``Auger Project''), 
   October 1994.

\end{thebibliography}
\end{document}